\documentclass[aps,floatfix,prl,nofootinbib,superscriptaddress,reprint,showpacs,10pt,preprintnumbers,longbibliography]{revtex4-1}
\usepackage[utf8]{inputenc}
\usepackage[pdftex]{graphicx}
\usepackage{float}
\usepackage{amsmath}
\usepackage{amssymb}
\usepackage{amsfonts}
\usepackage{dsfont}
\usepackage{array}
\usepackage{bm}
\usepackage{mathrsfs}
\usepackage{pifont}
\usepackage{multirow}
\usepackage{upgreek}
\usepackage[dvipsnames]{xcolor}
\usepackage[pdftex,
            pdftitle={The rho-resonance with physical pion mass from Nf=2 lattice QCD},
            pdfauthor={Matthias Fischer, Bartosz Kostrzewa, Maxim Mai, Marcus Petschlies, Ferenc Pittler, Martin Ueding, Carsten Urbach},
            bookmarks,
            colorlinks,
            linkcolor=myblue,
            citecolor=mymagenta,
            menucolor=black,
            urlcolor=myblue,
            plainpages=false,
            pdfpagelabels,
            hypertexnames=false]{hyperref}
\usepackage{verbatim}
\usepackage{slashed}
\usepackage{ulem}

\usepackage[capitalise]{cleveref}
\crefformat{figure}{#2Fig.~#1#3}
\crefmultiformat{figure}{Figs.~#2#1#3}{ and~#2#1#3}{, #2#1#3}{ and~#2#1#3}

\usepackage{ucs}
\usepackage[utf8]{inputenc}
\usepackage{cancel}

\definecolor{mymagenta}{RGB}{200, 0, 100}
\definecolor{myblue}{RGB}{45, 48, 146}

\graphicspath{{pics/}}

\begin{document}
\title{The $\rho$\,-\,resonance with physical pion mass from $N_f=2$ lattice QCD}
\author{Matthias Fischer}
\affiliation{Helmholtz Institut f\"ur Strahlen- und Kernphysik, University of Bonn, Nussallee 14-16, 53115 Bonn, Germany}
\affiliation{Bethe Center for Theoretical Physics, University of Bonn, Nussallee 12, 53115 Bonn, Germany}
\author{Bartosz Kostrzewa}
\affiliation{High Performance Computing \& Analytics Lab, Digital Science Center, Institut f\"ur Informatik, Endenicher Allee 19A, 53115 Bonn, Germany}
\author{Maxim Mai}
\affiliation{The George Washington University, Washington, DC 20052, USA}
\author{Marcus Petschlies}
\affiliation{Helmholtz Institut f\"ur Strahlen- und Kernphysik, University of Bonn, Nussallee 14-16, 53115 Bonn, Germany}
\affiliation{Bethe Center for Theoretical Physics, University of Bonn, Nussallee 12, 53115 Bonn, Germany}
\author{Ferenc Pittler}
\affiliation{Computation-based Science and Technology Research
  Center,The Cyprus Institute, 20 Kavafi Str., Nicosia 2121, Cyprus}
\author{Martin Ueding}
\affiliation{Helmholtz Institut f\"ur Strahlen- und Kernphysik, University of Bonn, Nussallee 14-16, 53115 Bonn, Germany}
\affiliation{Bethe Center for Theoretical Physics, University of Bonn, Nussallee 12, 53115 Bonn, Germany}
\author{Carsten Urbach}
\affiliation{Helmholtz Institut f\"ur Strahlen- und Kernphysik, University of Bonn, Nussallee 14-16, 53115 Bonn, Germany}
\affiliation{Bethe Center for Theoretical Physics, University of Bonn, Nussallee 12, 53115 Bonn, Germany}
\author{Markus Werner}
\affiliation{Helmholtz Institut f\"ur Strahlen- und Kernphysik, University of Bonn, Nussallee 14-16, 53115 Bonn, Germany}
\affiliation{Bethe Center for Theoretical Physics, University of Bonn, Nussallee 12, 53115 Bonn, Germany}

\collaboration{Extended Twisted Mass Collaboration}
\date{\today}
\begin{abstract}
  We present the first-ever lattice computation of
  $\pi\pi$--scattering in the $I=1$ channel with $N_f=2$ dynamical quark flavours obtained
  including an ensemble with physical value of the pion mass.
  Employing a global fit to data at three values of the pion mass,
  we determine the universal parameters of the $\rho$-resonance.
  We carefully investigate systematic uncertainties by determining
  energy eigenvalues using different methods and by comparing inverse
  amplitude method and Breit-Wigner type parametrizations. 
  Overall, we find mass $M_\rho=786(20)\ \mathrm{MeV}$ and
  width $\Gamma_\rho = 180(6)\ \mathrm{MeV}$, including statistical
  and systematic uncertainties. In stark disagreement with the previous
  $N_f=2$ extrapolations from higher than physical pion mass results, our mass
  value is in good agreement with experiment, while the width is slightly
  too high.
\end{abstract}

\maketitle

\textit{Introduction.}---Quantum Chromodynamics (QCD) -- the theory of
strong interactions -- gives rise to a fascinating plethora of
hadronic states: mesons and baryons. A theoretical understanding of
these states from first principles requires a
non-perturbative method, as provided by lattice QCD. As most of the
hadrons are not stable under the strong interaction and decay, such an
investigation must include resonance and interaction parameters.

\begin{figure}
  \includegraphics[width=\linewidth, page=2]{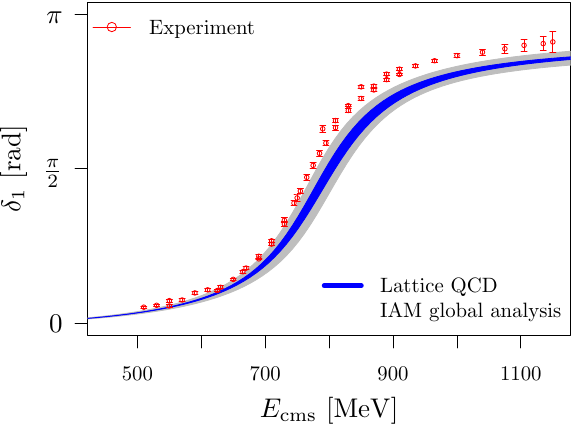}
  \caption{\label{plot:phaseshift-phys}
    $\rho$-meson $p$-wave phase-shift $\delta_1$ at the physical
    point. We compare experimental
    data~\cite{Estabrooks:1974vu,Protopopescu:1973sh} to our
    prediction from three global IAM fits, see text. The blue band
    visualises the statistical and fitting uncertainties, the gray
    band includes in addition the estimated lattice artefacts added in
    quadrature.}
\end{figure}

One very prominent such state is the so-called $\rho$-resonance
($\rho(770)$), which decays predominantly in a $p$-wave into two pions
with isospin $I=1$. It is experimentally observed as a peak in
cross-sections at an energy of $M_\rho\sim775\ \mathrm{MeV}$ with width
$\Gamma_\rho \sim 150\ \mathrm{MeV}$~\cite{PhysRevD.98.030001}. The
corresponding $p$-wave phase-shift curve is a prime example for a
resonance phase-shift, as can be seen from \cref{plot:phaseshift-phys},
where the experimentally measured phase-shift
$\delta_1$~\cite{Estabrooks:1974vu,Protopopescu:1973sh} is depicted
as a function of the center-of-mass energy.
Via vector meson dominance, the $\rho$ plays a fundamental role in our
theoretical understanding of many processes~\cite{Meissner:1987ge}
and, since it is well investigated experimentally, it represents a benchmark
resonance state for lattice QCD simulations.

The $\rho$-resonance has been investigated in lattice QCD
previously~\cite{Feng:2010es,Lang:2011mn,Aoki:2011yj,Dudek:2012xn,Bali:2015gji,
Wilson:2015dqa,Fu:2016itp,Guo:2016zos,Alexandrou:2017mpi,Andersen:2018mau},
most recently in Ref.~\cite{Werner:2019hxc} including a continuum
and chiral extrapolation. The latter was needed because the lattice
simulations were performed at unphysically large values of the pion
mass. One of the interesting conclusions from
Ref.~\cite{Werner:2019hxc} is that the chiral extrapolation is
difficult, even though there is guidance from effective field
theories. The main reason for this was the lack of ensembles
with sufficiently light pion mass, the lightest being at $230\ \mathrm{MeV}$.
In addition, there is an ongoing discussion about the origin of the
surprisingly large difference between the $\rho$-resonance parameters
obtained in $N_f=2$ and $N_f=2+1(+1)$ flavour
QCD~\cite{Molina:2016qnm,Hu:2017wli,Mai:2019pqr,Molina:2020qpw}.

With this letter we fill the gap to realistic pion mass values by including an ensemble directly at the physical point for the first time.
We present $\rho$-resonance parameters determined on three $N_f=2$ lattice QCD ensembles generated by the Extended Twisted Mass Collaboration (ETMC)~\cite{Abdel-Rehim:2015pwa} with pion masses between $132$ and $340\ \mathrm{MeV}$.

These allow us to interpolate rather than extrapolate to the physical pion mass and, therefore, to directly compare to experiment.
Moreover, by comparing to other lattice investigations we can shed new light on the question of the importance of the $\bar KK$ threshold for the $\rho$-resonance phase-shift.

The main result of this letter is summarised in \cref{plot:phaseshift-phys},
where, in addition to the experimental data
for the phase-shift, the blue band shows our interpolation for
the phase-shift at the physical pion mass value. The width of the band
represents fitting and statistical uncertainties. The gray band
includes in addition an estimate of lattice artefacts.

\textit{Lattice Computation.}---The results for the $\rho$-resonance properties presented in this letter are based on gauge configurations generated by the ETMC at a single value of the lattice spacing, $a = 0.0914(15)\ \mathrm{fm}$~\cite{Abdel-Rehim:2015pwa}.
These employ the Iwasaki gauge action~\cite{Iwasaki:1985we} and two dynamical mass-degenerate flavours of Wilson twisted mass clover fermions at maximal twist~\cite{Frezzotti:2000nk,Frezzotti:2003xj}.
With this action, physical quantities are $\mathcal{O}(a)$ improved~\cite{Frezzotti:2003ni} such that discretisation effects only appear at order $a^2$ in the lattice spacing.
The three ensembles considered for this letter are compiled in \cref{tab:ensembles} together with the lattice volume $(L/a)^3\times T/a$, the number of configurations $N_\mathrm{conf}$ and the pion mass value in physical units.
For more details we refer to Ref.~\cite{Abdel-Rehim:2015pwa}.

\begin{figure*}
  \includegraphics[width=0.98\textwidth]{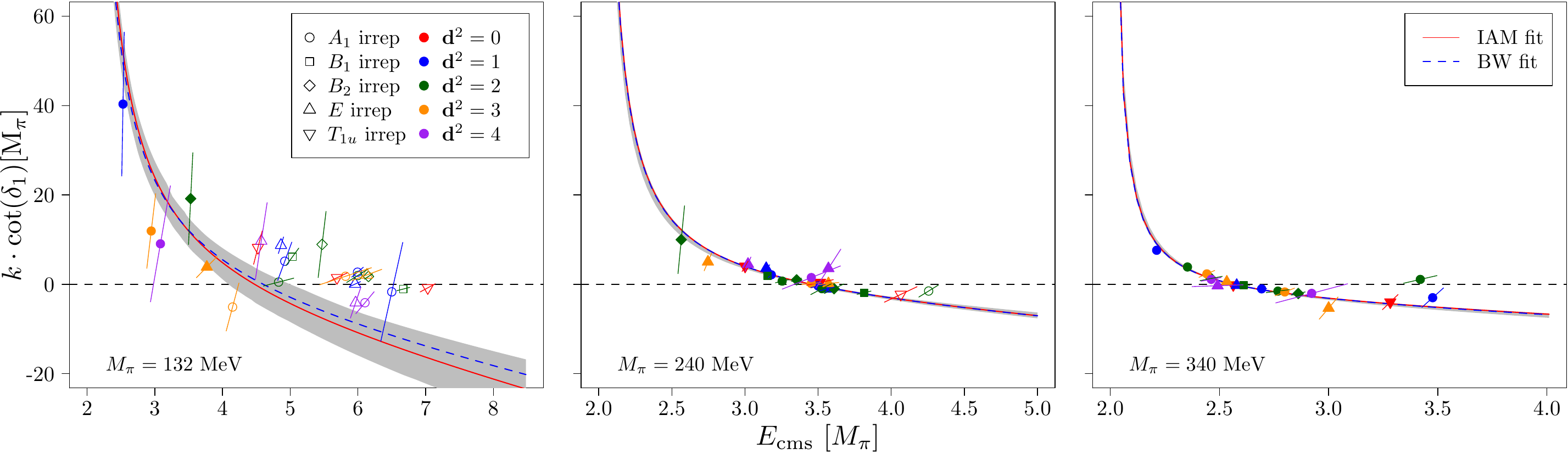}
  \caption{\label{plot:phaseshift-fits}
    $k\cot(\delta_1)$ as a function of $E_\mathrm{cms}$, both in units
    of $M_\pi$,
    obtained from the fits to energy eigenvalues of analysis \textbf{A1} for the three
    individual ensembles separately (from left to right: cA2.09.48, cA2.30.48
    and cA2.60.32). Full red and blue dashed lines
    show results of IAM- and BW-based fits, respectively.
    Data points are added for illustration purposes only. For better
    legibility we have dropped a few points with absolute error larger
    than 20 and relative error larger than 75\%.
  }
\end{figure*}

\begin{table}
  \centering
  \begin{tabular*}{.49\textwidth}{@{\extracolsep{\fill}}lcccc}
    \hline\hline
    Ens & $(L/a)^3\times T/a$ & $N_\mathrm{conf}$ & $M_\pi\ [\mathrm{MeV}]$ \\
    \hline\hline
    $cA2.09.48$ & $48^3\times96$ & $1485$ & $132$ \\
    $cA2.30.48$ & $48^3\times96$ & $343$ & $240$ \\
    $cA2.60.32$ & $32^3\times64$ & $334$ & $340$ \\
    \hline\hline
  \end{tabular*}
  \caption{Parameters and pion mass values of the ensembles used. All
    ensembles have $\beta=2.10$ and clover coefficient
    $c_\mathrm{SW}=1.57551$ in common.}
  \label{tab:ensembles}
\end{table}

On these ensembles, we compute center-of-mass energy levels
$E^\Gamma_\mathrm{cms}$ for irreducible representations (irreps) of
the lattice rotational symmetry group $\Gamma$.
We follow the procedure detailed in Ref.~\cite{Werner:2019hxc}
and compute Euclidean correlation matrices
$\mathcal{C}_{\Gamma,\mathbf{p}^2}$
\begin{equation}
  \mathcal{C}_{\Gamma,\mathbf{p}^2}(t)\ =\ \langle
  \mathcal{O}_{\Gamma, \mathbf{p}}(t'+t)\cdot
  \mathcal{O}_{\Gamma, \mathbf{p}}(t')^\dagger\rangle
\end{equation}
averaging over all equivalent momenta $\mathbf{p}$. The operators
$\mathcal{O}_{\Gamma,\mathbf{p}} = (\mathcal{O}_\Gamma^1, \mathcal{O}_\Gamma^2,
\ldots)^t$ are chosen to project to irrep $\Gamma$ for total squared momentum
$\mathbf{p}^2$. The list of irreps $\Gamma$ considered is $T_{1u}, A_1,
E, B_1, B_2$ up to $\mathbf{d}^2=4$ with
$\mathbf{p}=2\pi\mathbf{d}/L$.

The basis operators used to construct the operators
$\mathcal{O}_{\Gamma,\mathbf{P}}$ are two pion and single
vector meson operators
\begin{equation}
  \begin{split}
    \mathcal{O}_{\pi^+\pi^-}(x,y)\ &=\ \bar d\, i\gamma_5\, u(x)\ \bar
    u\, i\gamma_5\, d(y)\,,\\
    \mathcal{O}_\rho(x)\ &=\ \frac{1}{\sqrt{2}}(\bar u \Gamma^\rho u(x) - \bar d
    \Gamma^\rho d(x))\\
  \end{split}
\end{equation}
with $\Gamma^\rho\in\{i\gamma_i, \gamma_0\gamma_i\}$.

We apply the generalised eigenvalue method (GEVM), i.e., solve the
generalised eigenvalue
problem~\cite{Michael:1982gb,Luscher:1990ck}
for eigenvalues $\lambda^{(n)}(t)$
and eigenvectors $\eta^{(n)}$, where $n$ labels the contributing states. Energy levels of these can be determined from the
exponential fall-off of $\lambda^{(n)}(t)$ at large $t$.
In addition, we apply the so-called Prony generalised eigenvalue
method (PGEVM) in form of a matrix pencil on top of the
GEVM~\cite{Fischer:2020aaa} to reduce excited state contaminations.

The confidence in our energy eigenvalue extractions is increased by employing the following three methods:
\begin{itemize}
\item[\textbf{A1}:] direct fit to each $\lambda^{(n)}(t)$ using a fit range chosen by eye.
\item[\textbf{A2}:] direct fit to each $\lambda^{(n)}(t)$ using the fit range which yields the fit with the best $p$-value.
\item[\textbf{A3}:] fit to the principal correlator of the
  PGEVM 
  obtained from $\lambda^{(n)}(t)$~\cite{Fischer:2020aaa} using a fit
  range chosen by eye.
\end{itemize}
Energy eigenvalues for which the three methods do not yield consistent results are discarded.
To account for residual deviations, we perform the resonance parameter determinations based on energy eigenvalues obtained using each method and take the maximal difference between the resulting parameters as a systematic uncertainty.

When considering multi-particle operators with periodic boundary conditions, the corresponding correlation functions are polluted by contributions from the so-called thermal states and there exist several methods to reduce or remove these.
However, in Ref.~\cite{Werner:2019hxc} we have shown that in the $I=1$ channel, the extracted energies agree within errors with or without thermal state subtraction if the fit-range is chosen carefully.
We have checked that this is the case also here and thus use energy levels extracted without thermal state subtraction.
Like in Ref.~\cite{Werner:2019hxc}, we use so-called stochastic Laplacian Heaviside smearing~\cite{Morningstar:2011ka} with algorithmic parameters identical to Ref.~\cite{Dimopoulos:2018xkm}.
For the determination of the pion decay constant on the same gauge configurations, we also employ local time slice sources and the so-called one-end-trick, for details see Ref.~\cite{Boucaud:2008xu}.

\renewcommand{\arraystretch}{1.2}
\begin{table}[b]
  \centering
  \normalsize
  \begin{tabular*}{.499\textwidth}{@{\extracolsep{\fill}}llllll}
    \hline\hline
    Ens & $\bar l_{12}$ & $af_0$ & $a\Lambda_4$ &
    $\chi^2_{\rm dof}$ & dof \\
    \hline\hline
    cA2.09.48 & $-11.0^{+2.5}_{-2.8}$ & $0.060^{+0.002}_{-0.003}$ & $0.088^{+0.284}_{-0.067}$ & $1.6$ & $4$ \\
    cA2.30.48 &$-6.8^{~+0.8}_{~-0.4}$ & $0.065^{+0.002}_{-0.004}$ & $0.138^{+0.167}_{-0.039}$ & $0.9$& $25$\\
    cA2.60.32 &$-6.3^{~+0.2}_{~-0.8}$ & $0.063^{+0.004}_{-0.001}$ & $0.041^{+0.041}_{-0.104}$ & $0.8$ & $23$\\
    \hline
    all       &$-5.3^{~+0.0}_{~-0.1}$ & $0.057^{+0.000}_{-0.000}$ & $0.543^{+0.010}_{-0.006}$ & $1.1$ & $58$\\
    \hline\hline
  \end{tabular*}
  \caption{\label{tab:fits}
    Results of IAM-based correlated fits to single ensembles and
    global set of eigenvalues, including statistical uncertainties
    determined from re-sampling. Fits are based on energy levels
    estimated with method \textbf{A1}, for \textbf{A2-3} see
    \cite{SupplMat}.}
\end{table}

\textit{Phase-shift determination.}---The discrete and real
valued lattice energy levels $E_{\mathrm{cms}}^\Gamma$ are mapped to
the infinite volume scattering quantities using Lüscher's
method~\cite{Luscher:1986pf,Luscher:1990ux,Luscher:1990ck}. In case
of the $\rho$-meson and under the assumption that higher partial waves
can be neglected, the $p$-wave phase-shift
$\delta_1$ is related to the energy levels via
\begin{equation}
  \label{eq:luescher_formula}
  \cot \delta_1 = M^{\Gamma}(k^2)\,,
\end{equation}
where $M^\Gamma$ is an algebraically known matrix
function~\cite{Werner:2019hxc,Bulava:2016mks,Morningstar:2017spu} of
the lattice scattering momentum ${k^2(E^2_\mathrm{cms}) =
  E_\mathrm{cms}^2/4 - M_\pi^2}$ and the pion mass, $M_\pi$. Note that \cref{eq:luescher_formula}
is valid below inelastic threshold ($4M_\pi$) only.
This represents a limitation in particular for the ensemble cA2.09.48,
where only five energy levels lie below this threshold for our $L$-value.
In a more general sense this also implies that independently of the
number of points below threshold, the resonance region of the $\rho$-meson
can never be mapped out using Lüscher's method only, because
$4M^\mathrm{phys}_\pi< M_\rho$.

Given only discrete values of $E_\mathrm{cms}$, one needs to parameterize
the scattering amplitude as a function of a continuous $E_\mathrm{cms}$. One
example for such a parametrization is a simple Breit-Wigner (BW) form
\begin{equation}
  \label{eq:BWphase}
  \tan\delta^\mathrm{BW}_1(s) = \frac{g_{\rho\pi\pi}^2}{6\pi}
  \frac{k^3(s)}{\sqrt{s}\,(M_\rho^2 - s)}\,,
\end{equation}
with $M_\rho$ the $\rho$-resonance mass, $g_{\rho\pi\pi}$ the
$\rho-\pi\pi$ coupling and $s$ the center-of-mass energy squared.

Supplementary to the experimental measurements, additional information
about the dynamics of the $\pi\pi$ system resides in the pion-mass dependence,
which can be explored with lattice calculations.
Being in the unique position of having data at the physical, as well as heavier than physical pion mass values, we use the IAM parametrization of the scattering amplitude ~\cite{Truong:1988zp,Dobado:1996ps,GomezNicola:2007qj,GomezNicola:2007qj}.
This approach preserves unitarity exactly, has the correct pion mass
dependence up to next-to-leading order (NLO) in chiral perturbation
theory~\cite{Gasser:1983yg,Gasser:1984gg} and fulfills further
non-perturbative constraints on the chiral trajectory~\cite{Bruns:2017gix}.

In IAM, the phase-shift $\delta_1$ is parameterized as (for more
details see Ref.~\cite{Mai:2019pqr})
\begin{equation}
  \label{eq:IAMphase}
  \cot\,\delta_1^\mathrm{IAM}(s)=
  \frac{\sqrt{s}}{2k}
  \Bigg(\frac{T_2(s)-\bar T_4(s)}{(T_2(s))^2}
  -16\pi\operatorname{Re}{J(s)}\Bigg)\,,
\end{equation}
where $T_2$ denotes the leading chiral order amplitude and $\bar T_4$
the NLO one without $s$-channel loop diagrams. The two-meson
loop in dimensional regularization is denoted by $J(s)$.
The corresponding amplitude is regularization scale independent and
depends on one combination of low-energy constants
(LECs)~\cite{Gasser:1983yg} $\bar l_{12}:=\bar l_1-\bar l_2$ as well as the pion
decay-constant in the chiral limit ($f_0$). Note that both $T_2$ and $T_4$
are expressed in terms of $M_\pi^2/(4\pi f_0)^2$.

The expressions~\eqref{eq:BWphase} and~\eqref{eq:IAMphase} are
fitted directly to the energy eigenvalues using
\cref{eq:luescher_formula} without computing the phase-shifts as
an intermediate quantity, or performing any scale setting.
Fit parameters are the BW parameters or
aforementioned LECs, depending on the considered fit form. In both
cases $M_\pi$ is fitted to the lattice values including finite size
corrections as described in Ref.~\cite{Colangelo:2005gd}.
Additionally, in the case of the IAM we
also fit $f_\pi$ with respect to a further constant, $\Lambda_4$,
related to the NLO LEC $\bar l_4$~\cite{Gasser:1983yg}. For details see
Ref.~\cite{SupplMat}. In all fits we take full account of correlations
and compute statistical uncertainties using the bootstrap. All bare
data is publicly available in a data repository~\cite{datarepo}.

\begin{table}[b]
  \centering
  \normalsize
  \begin{tabular*}{.49\textwidth}{@{\extracolsep{\fill}}llll}
    \hline\hline
    Ens & Method & $\operatorname{Re} E_\rho$~[MeV] & $\operatorname{Im} E_\rho$~[MeV]\\
    \hline\hline
    cA2.09.48
    & IAM & $587.3^{+65.7}_{-49.1}$ & $28.8^{+14.1}_{-8.7}$ \\
    & BW  & $603.1^{+228.2}_{-86.9}$ & $34.2^{+171.9}_{-24.3}$ \\
    cA2.30.48
    & IAM & $821.0^{+0.0}_{-11.8}$ & $48.0^{+5.0}_{-4.1}$ \\
    & BW  & $821.0^{+0.0}_{-11.8}$ & $48.0^{+5.0}_{-4.1}$ \\
    cA2.60.32
    & IAM & $868.0^{+1.7}_{-5.4}$ & $24.1^{+0.3}_{-2.7}$ \\
    & BW  & $868.0^{+1.7}_{-5.8}$ & $24.1^{+0.3}_{-2.7}$ \\
    \hline
    all & global IAM & $786.8^{+0.1}_{-5.2}$ & $90.1^{+0.0}_{-2.0}$ \\
    \hline\hline
  \end{tabular*}
  \caption{\label{tab:poles}
    Pole positions $E_\rho$ determined using IAM and BW
    parametrizations for the different lattice ensembles with method
    \textbf{A1} to estimate the energy levels. Last row
    shows the extrapolation of the global IAM fit to the physical
    point.}
\end{table}

\begin{figure}[t]
  \centering
  \includegraphics[width=\linewidth, trim=0 0.8cm 0 0]{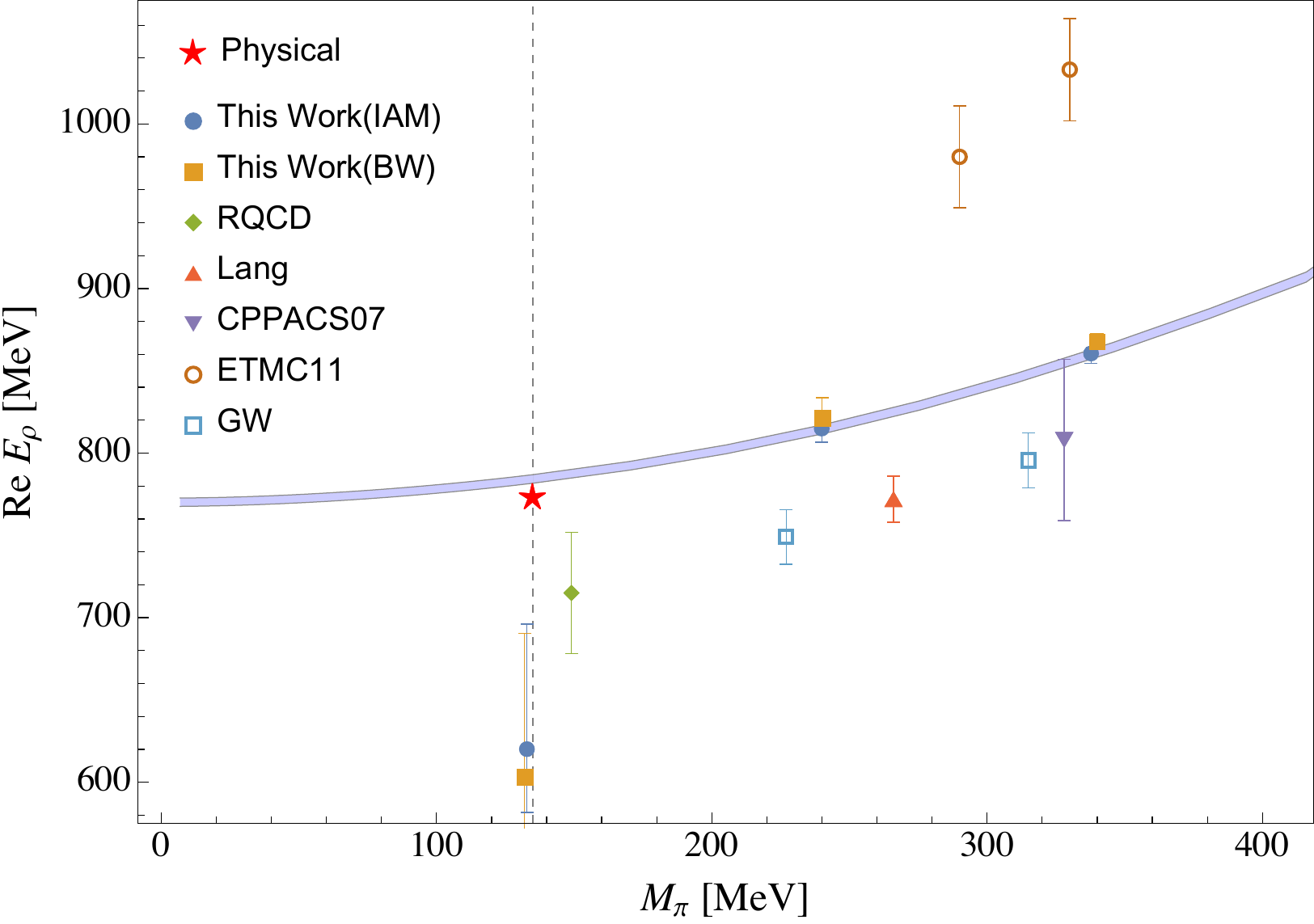}
  \caption{
    Compilation of results of this and other
    works~\cite{Bali:2015gji,Aoki2013,Lang:2011mn,Feng:2010es} on
    $M_\rho$ as function of the pion mass. The indicated error bars
    combine systematic and statistical uncertainties. The blue shaded
    band shows the pion mass dependence of our global fit. We quote
    the PDG central value~\cite{PhysRevD.98.030001} by the red star at
    135~MeV (dashed vertical line) for comparison.}
  \label{plot:rho-mass}
\end{figure}

\textit{Results.}---Here we present and
discuss mainly the results obtained with method \textbf{A1} to
estimate energy levels if not mentioned otherwise. For methods
\textbf{A2-3} see \cite{SupplMat}. In \cref{plot:phaseshift-fits}
we show $k\cot(\delta_1)$ as a function of the
center-of-mass energy $E_\mathrm{cms}$ both in units of the pion mass for
the three ensembles separately. The solid red lines with $1\sigma$
error band correspond to the best fits of \cref{eq:IAMphase} directly
to the energy levels on each ensemble separately, the blue dashed
lines to Breit-Wigner fits \cref{eq:BWphase}.
The data points with slanted error bars indicating the
correlation between $\delta_1$ and $E_\mathrm{cms}$ are generated using
\cref{eq:luescher_formula} for illustration purposes only. Filled
symbols correspond to data points with $E_\mathrm{cms}/M_\pi\leq 4$,
which are included in the fits, open symbols to the rest.

\begin{figure}[t]
  \centering
  \includegraphics[width=\linewidth, trim=0 0.8cm 0 0]{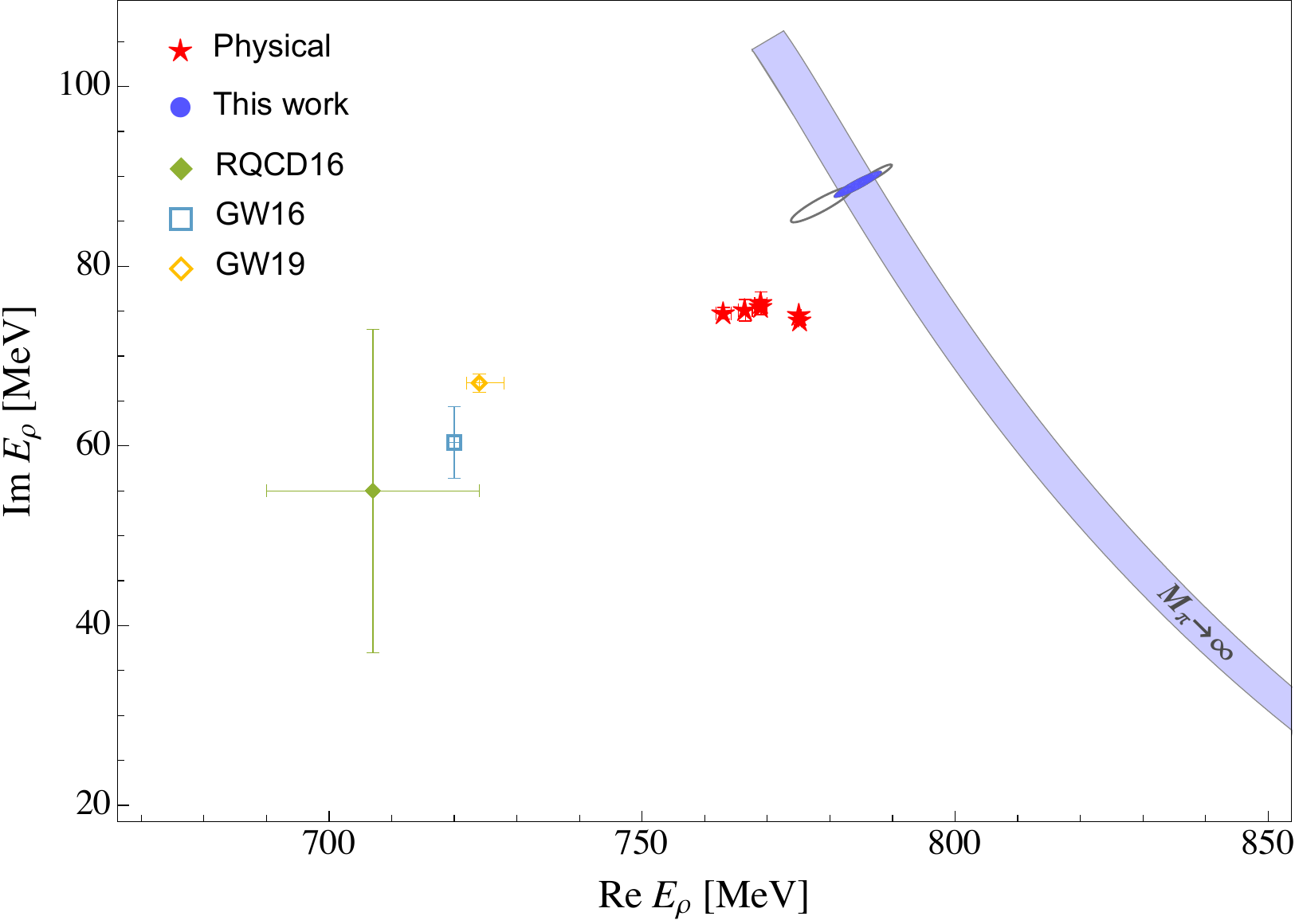}
  \caption{
    The complex pole position of the $\rho$-meson. The blue ellipse shows
    the $1\sigma$ boundary of the pole positions (global fit {\bf A1}) at the
    physical point. Corresponding ellipses for global fits to methods {\bf A2-3}
    are depicted as gray empty ellipses. Blue shaded band shows the pion mass dependence of
    the pole position with corresponding $1\sigma$-band. PDG results~\cite{PhysRevD.98.030001} and
    those of earlier lattice calculations~\cite{Molina:2016qnm,Bali:2015gji,Mai:2019pqr} are quoted for comparison.}
  \label{plot:rho-pole}
\end{figure}

The results of our fits and the corresponding
$\chi^2_{\rm dof}$ values are compiled in \cref{tab:fits}.
The complex $\rho$-resonance pole position $E_\rho=M_\rho +
\mathrm{i}\Gamma_\rho/2$ is given in \cref{tab:poles}, the last row of
which gives the pole position at the physical point, which we define using
$M_\pi/f_\pi=135/130.41$ and $M_\pi^\mathrm{phys}=135\ \mathrm{MeV}$ as
input~\cite{SupplMat}, resulting in $a=0.0919(1)\ \mathrm{fm}$
(statistical error only) well compatible with
Ref.~\cite{Abdel-Rehim:2015pwa}. The
results of the single (IAM, BW) and global (IAM) fits
are depicted in \cref{plot:rho-mass} together with the physical
result~\cite{PhysRevD.98.030001} and previous $N_f=2$ lattice
determinations~\cite{Bali:2015gji,Aoki2013,Lang:2011mn,Feng:2010es}.
The complex  pole positions of our global fit are
visualised in \cref{plot:rho-pole}. The solid $1\sigma$ error band
represents $E_\rho$ determined by the global IAM fit as a function of
the pion mass (from larger pion mass in the bottom right to smaller
pion mass in the top left corner of the plot), the blue ellipse indicates the
corresponding pole position at the physical point.
The two grey ellipses correspond to pole positions from global IAM fits \textbf{A2} and
\textbf{A3}. In addition, we show
PDG values~\cite{PhysRevD.98.030001}, indicating the
variation in the phenomenological extractions, and results
of chiral extrapolations of previous heavier pion mass $N_f=2$ lattice
determinations~\cite{Molina:2016qnm,Bali:2015gji,Mai:2019pqr}. 

Finally, in \cref{plot:phaseshift-phys} we show the experimental
phase-shift data~\cite{Estabrooks:1974vu,Protopopescu:1973sh} as a
function of $E_\mathrm{cms}$ and compare to our global IAM fit
prediction for the physical pion mass value. For the
latter we plot in blue the envelope area of all the error bands of the three
global IAM analyses \textbf{A1-3}, thus, visualising statistical and
IAM uncertainties. The grey band includes our estimate of the lattice
artefacts added in quadrature.

\textit{Discussion.}---First, we observe very good agreement
between BW and IAM fits on the three ensembles separately, as can be
see in \cref{tab:fits,tab:poles}, with smaller errors
in the IAM fits. This is even the case on the physical point
ensemble, however, with much too low pole mass and width. The latter
can be attributed to only five points being included in the fit, with much
of the curvature triggered by a single data point, because all the
points are in a region where $\delta_1$ is close to zero.

This emphasises the importance of including ensembles with larger than
physical pion mass value in the analysis: only on those ensembles can
the Lüscher method cover the resonance region fully.

Interestingly, if one were to ignore the inelastic threshold on the
physical point ensemble and use \cref{eq:luescher_formula} to obtain
values for $\delta_1$, the such obtained phase-shift points compare
reasonably well with the experimental data~\cite{SupplMat}. This is due to the fact that
$\rho\to\pi\pi$ is almost elastic up to $1\ \mathrm{GeV}$, which is
actually also the assumption to obtain the experimental phase-shift
points (see also Ref.~\cite{Gasser:1990bv}).

The three global IAM analyses based on energy determinations
\textbf{A1-3} agree very well with each other (see Tab.~1 in
\cite{SupplMat}) and we conclude that they lead to consistent chiral
extrapolations. Though the physical point ensemble cA2.09.48 might not, due to the few
energy levels, add much information to the global fit, it is very
important, because it anchors the fit at slightly below the physical
pion mass value. This turns our determination of the $\rho$-resonance
properties into an interpolation, making it much more reliable.

Below, we take the maximum of the $\pm$ statistical errors as
our statistical uncertainty, the maximal deviation between the
three analyses \textbf{A1-3} as a systematic uncertainty.
In addition we assign a generic $2.5$\% uncertainty to (undetermined)
discretisation artefacts, which are generically of order $a^2\Lambda_\mathrm{QCD}^2$.
Therefore, we quote as our final result the results from the global
IAM fit with analysis method \textbf{A1}, reading
\begin{equation}
  \label{eq:final}
  \begin{split}
    M_\rho\ &=\ 786~(5)_\mathrm{stat}~(1)_\mathrm{sys}~(19)_\mathrm{lat}\ \mathrm{MeV}\,,\\
    \Gamma_\rho\ &=\ 180~(4)_\mathrm{stat}~(1)_\mathrm{sys}~(4)_\mathrm{lat}\ \ \, \mathrm{MeV}\,,\\
    \bar{l}_{12} \ &=\ -5.27~(8)_\mathrm{stat}~(3)_\mathrm{sys}~(13)_\mathrm{lat}\,,\\
    \bar{l}_4\ &=\ +4.31~(4)_\mathrm{stat}~(2)_\mathrm{sys}~(10)_\mathrm{lat}\,,\\
    f_0\ &=\ 86.46~(3)_\mathrm{stat}~(3)_\mathrm{sys}~(2)_\mathrm{lat}\,.\\
  \end{split}
\end{equation}
Note that our parametrically estimated lattice artefacts
are the largest source of uncertainty in our results.

Compared to other analyses of $N_f=2$ lattice QCD data for the
$\rho$-meson, our results are much closer to the values quoted in the
PDG. In particular, we obtain a slightly larger value for the
pole mass, which is in contrast to the claim made in
Ref.~\cite{Molina:2016qnm} that the missing $K\bar{K}$ channel pushes
 this value downwards in $N_f=2$ QCD.
From our results we can only conclude that the influence of the
missing strange quark in our ensembles is marginal. There are other
effects, most likely lattice artefacts, which are able to explain the
findings in Ref.~\cite{Molina:2016qnm}, see also recent analyses~\cite{Hu:2017wli,Mai:2019pqr,Molina:2020qpw}.
Note also that a different scale setting procedure (see Ref.~\cite{Abdel-Rehim:2015pwa}) would not change our results sufficiently.

Compared to Ref.~\cite{Werner:2019hxc} with $N_f=2+1+1$ dynamical
quark flavours, we have presented in this letter a significantly
better controlled chiral extrapolation thanks to the included physical
point ensemble. While the pole mass is similarly close to the
experimental value, our width is larger than the physical one, whereas
in Ref.~\cite{Werner:2019hxc} a lower value was found. A final
estimate will require a continuum extrapolation at the physical pion
mass value.

\textit{Conclusion.}---We have presented a lattice QCD analysis of the
$\rho$-resonance including an ensemble with slightly lower than
physical pion mass value for the first time. This allows us to
interpolate to the physical point using the inverse amplitude method
including the pion mass dependence up to NLO in the chiral expansion.

With all our uncertainties added in quadrature, our results for the $\rho$-meson
mass and width read
\[
M_\rho\ =\ 786(20)\ \mathrm{MeV}\,,\quad \Gamma_\rho\ =\ 180(6)\ \mathrm{MeV}\,.
\]
While $M_\rho$ agrees well with the PDG value, the width is too large by $20$\%.
The low energy constants are in very good agreement with the
corresponding FLAG lattice averages~\cite{Aoki:2019cca}. This is not
necessarily expected, since the IAM resums higher order effects due to
unitarisation.

Eventually, this computation needs to be repeated with $N_f=2+1(+1)$
dynamical quark flavours, several values of the lattice spacing and
physical point ensembles included. Particular emphasis should also be
on different spatial volumes at the physical point.

\begin{acknowledgments}
  \textit{Acknowledgments.}---
  We thank all members of the ETMC for the most enjoyable
  collaboration. The authors gratefully acknowledge the Gauss Centre for Supercomputing
  e.V. (www.gauss-centre.eu) for funding this project by providing
  computing time on the GCS Supercomputer JUQUEEN~\cite{juqueen} and the
  John von Neumann Institute for Computing (NIC) for computing time
  provided on the supercomputers JURECA~\cite{jureca} and JUWELS~\cite{juwels} at Jülich
  Supercomputing Centre (JSC). We thank Ulf-G.~Meißner for useful
  comments on the manuscript.
  This project was funded in part by the DFG as a project in the
  Sino-German CRC110. FP acknowledges financial support from the
  Cyprus Research and Innovation Foundation under project “NextQCD”,
  contract no. EXCELLENCE/0918/0129.
  The open source software packages tmLQCD~\cite{Jansen:2009xp,Abdel-Rehim:2013wba,Deuzeman:2013xaa},
  Lemon~\cite{Deuzeman:2011wz},
  QUDA~\cite{Clark:2009wm,Babich:2011np,Clark:2016rdz}, the hadron
  package~\cite{hadron:2020} and R~\cite{R:2005} have been used.
  MM thanks R.~Brett and A.~Alexandru for assistance in implementation of FinVol routines.
\end{acknowledgments}

\bibliographystyle{h-physrev5}

\onecolumngrid
\newpage
\begin{appendix}

  \begin{center}
    \Large Supplemental Material
  \end{center}

\section{Fit strategy and results}

We employ two distinct parametrizations of $\cot\delta_1$ as
discussed in the main part of the article. The first one, of
Breit-Wigner type, contains two parameters $M_\rho$ and
$g_{\rho\pi\pi}$, which are determined in a fit to energy eigenvalues
using Lüscher's
method~\cite{Luscher:1986pf,Luscher:1990ck,Luscher:1990ck}. The width
$\Gamma_\rho$ is related to $M_\rho$ and $g_{\rho\pi\pi}$ via
\[
\Gamma_\rho\ =\ \frac{2}{3} \frac{g_{\rho\pi\pi}^2}{4\pi}
\frac{p^3(M_{\rho})}{M_{\rho}^2}\,,\qquad
p(M) = \sqrt{M^2/4 - M_\pi^2}\,.
\]
The fit is performed by minimizing the fully correlated $\chi^2$ with
respect to 3 free parameters, i.e., $\{M_\pi,M_\rho,g_{\rho\pi\pi}\}$. Since this parametrization does not
include the pion mass dependence, only single fits to each ensemble are
performed. The results for the dynamical parameters are given in
\cref{tab:fits} for method \textbf{A1} and in \cref{tab:IAMfits} for
all methods \textbf{A1-3}.

A second analysis is based on the so-called inverse amplitude method,
see the main text for the explicit parametrization.
As discussed there, this method currently allows for the most reliable extrapolation of $\pi\pi$ dynamics along the energy and pion mass directions.
However, to avoid biases, a careful discussion of its implementation with regard to scale setting is in order.
At the level of our fits, we completely avoid the need for scale
setting by expressing leading and next-to-leading chiral order
amplitudes ($T_2$ and $\bar T_4$) in terms of $M_\pi/f_0$. $f_0$ is
then related to $f_\pi$ -- for which we have data -- using the
corresponding NLO ChPT relation involving $\Lambda_4$.
In addition to $f_0$ and $\Lambda_4$, we are left with only one
combination of free parameters, namely $(\bar l_1-\bar l_2)$.
In individual fits the set of free parameters is given by
$\{aM_\pi,af_0,(\bar l_1-\bar l_2),a\Lambda_4\}$.
This allows to fit the energy eigenvalues together with the finite
volume pion mass ($M_\pi(L)$) and decay constant ($f_\pi(L)$) by
relating
\begin{align}
M_\pi(L)=M_\pi\left(1+\frac{1}{2}\xi g_1\right)
\quad\text{and}\quad
f_\pi(L, M_\pi^2)=f_0\left(1-2\xi g_1\right)\left((1-2\xi\log\left(\frac{M_\pi^2}{\Lambda_4^2}\right)\right)\,,
\end{align}
for $\xi=M_\pi^2/(4\pi f_0)^2$ and $\tilde g_1(x)=\sum_{n=1}^\infty
4m(n)/(\sqrt{n}x)K_1(\sqrt{n}x)$ with $K_1$ denoting the Bessel
function of the second kind. Multiplicities $m(n)$ and further
definitions can be found in Ref.~\cite{Colangelo:2005gd}. The fit is
performed by minimizing the fully correlated $\chi^2$, expressing
everything directly in lattice units. The global fit is performed in
a very similar way by simply extending the set of free parameters
to include two further pion masses, i.e.,
$\{aM_\pi^1,aM_\pi^2,aM_\pi^3,af_0,(\bar l_1- \bar l_2), a\Lambda_4 \}$.
Note that all other parameters are pion mass independent and, thus,
are common to all three ensembles. The results of all fits are given
 in \cref{tab:IAMfits} for all methods \textbf{A1-3} (see also
 \cref{tab:fits} in the main text).
Statistical errors on the fit parameters are estimated using the bootstrap.

\renewcommand{\arraystretch}{1.6}
\begin{table*}[h]
  \centering
  \tiny
  \begin{tabular}{|l|ll|l|ll||l|l|l|l|ll|}
  \hline
& \multicolumn{5}{c||}{BW} & \multicolumn{6}{c|}{IAM}\\
\cline{2-12}
Config.
& $g_{\rho\pi\pi}$    & $M_{\rho}~[M_\pi]$  & $\chi^2_{\rm dof}$ & \multicolumn{2}{c||}{$E_\rho~[M_\pi]$}
& $\bar l_{12}$ & $af_0$ & $a\Lambda_{4}$     & $\chi^2_{\rm dof}$ &  \multicolumn{2}{c|}{$E_\rho~[M_\pi]$}\\
\hline
\hline
cA2.09.48(\textbf{A1})
&$4.86^{+5.14}_{-1.81}$ & $4.602^{+2.513}_{-0.69}$ & $1.3$ & $4.569^{+1.729}_{-0.658}$&$-i\,0.259^{+1.302}_{-0.184}$
&$-11.03^{+2.46}_{-2.77}$ & $0.060^{+0.0022}_{-0.0025}$ & $0.0882^{+0.2844}_{-0.0671}$ & $1.6$ & $4.449^{+0.498}_{-0.372}$&$-i\,0.218^{+0.107}_{-0.066}$ \\
cA2.09.48(\textbf{A2})
&$19.99^{+0.01\star}_{-13.44}$ & $14.998^{+0.626}_{-9.707}$ & $2.1$ & $8.00^{+0.000}_{-2.836}$&$-i\,1.809^{+0.733}_{-1.242}$
&$-6.20^{+1.76}_{-3.12}$ & $0.0559^{+0.0001}_{-0.0001}$ & $0.9989^{+0.0011}_{-0.0029}$ & $3.1$ & $5.404^{+0.822}_{-0.908}$&$-i\,0.516^{+0.336}_{-0.255}$ \\
cA2.09.48(\textbf{A3})
&$19.98^{+0.01\star}_{-14.03}$ & $12.978^{+0.523}_{-8.587}$ & $1.7$ & $8.00^{+0.186}_{-3.676}$&$-i\,2.810^{+1.190}_{-2.457}$
&$-9.44^{+0.26}_{-5.30}$ & $0.0577^{+0.0007}_{-0.0004}$ & $0.3634^{+0.1027}_{-0.1075}$ & $2.8$ & $4.606^{+0.179}_{-0.877}$&$-i\,0.270^{+0.020}_{-0.152}$ \\
\hline
\hline
cA2.30.48(\textbf{A1})
&$5.76^{+0.36}_{-0.18}$ & $3.457^{+0.008}_{-0.053}$ & $0.9$ & $3.421^{+0.000}_{-0.049}$&$-i\,0.200^{+0.021}_{-0.017}$
&$-6.76^{+0.82}_{-0.46}$ & $0.0648^{+0.0016}_{-0.0042}$ & $0.1375^{+0.1666}_{-0.0390}$ & $0.9$ & $3.421^{+0.000}_{-0.049}$&$-i\,0.200^{+0.021}_{-0.017}$ \\
cA2.30.48(\textbf{A2})
&$5.49^{+0.37}_{-0.21}$ & $3.434^{+0.003}_{-0.042}$ & $0.5$ & $3.405^{+0.000}_{-0.039}$&$-i\,0.179^{+0.023}_{-0.016}$
&$-7.50^{+0.95}_{-0.64}$ & $0.0678^{+0.0020}_{-0.0045}$ & $0.0708^{+0.1113}_{-0.0246}$ & $0.5$ & $3.405^{+0.000}_{-0.039}$&$-i\,0.179^{+0.023}_{-0.016}$ \\
cA2.30.48(\textbf{A3})
&$5.24^{+0.78}_{-0.18}$ & $3.505^{+-0.066}_{-0.182}$ & $0.3$ & $3.480^{+0.070}_{-0.185}$&$-i\,0.172^{+0.036}_{-0.030}$
&$-8.25^{+2.06}_{-0.61}$ & $0.0728^{+0.0002}_{-0.0121}$ & $0.0221^{+0.2705}_{-0.0016}$ & $0.3$ & $3.48^{+-0.070}_{-0.185}$&$-i\,0.173^{+0.033}_{-0.029}$ \\
\hline
\hline
cA2.60.32(\textbf{A1})
&$5.91^{+0.08}_{-0.31}$ & $2.566^{+0.004}_{-0.017}$ & $0.8$ & $2.553^{+0.005}_{-0.017}$&$-i\,0.071^{+0.001}_{-0.008}$
&$-6.30^{+0.19}_{-0.78}$ & $0.0629^{+0.0040}_{-0.0012}$ & $0.0414^{+0.0414}_{-0.1038}$ & $0.8$ & $2.553^{+0.005}_{-0.016}$&$-i\,0.071^{+0.001}_{-0.008}$ \\
cA2.60.32(\textbf{A2})
&$5.93^{+0.07}_{-0.15}$ & $2.583^{+0.006}_{-0.008}$ & $1.4$ & $2.570^{+0.006}_{-0.008}$&$-i\,0.074^{+0.002}_{-0.004}$
&$-6.28^{+0.16}_{-0.36}$ & $0.0632^{+0.0018}_{-0.0009}$ & $0.0316^{+0.0316}_{-0.0516}$ & $1.5$ & $2.570^{+0.006}_{-0.008}$&$-i\,0.074^{+0.002}_{-0.004}$ \\
cA2.60.32(\textbf{A3})
&$6.00^{+0.05}_{-0.24}$ & $2.571^{+0.009}_{-0.038}$ & $0.6$ & $2.557^{+0.009}_{-0.036}$&$-i\,0.074^{+0.003}_{-0.009}$
&$-6.11^{+0.11}_{-0.54}$ & $0.0619^{+0.0024}_{-0.0014}$ & $0.0486^{+0.0486}_{-0.0725}$ & $0.6$ & $2.557^{+0.010}_{-0.035}$&$-i\,0.073^{+0.003}_{-0.009}$ \\
\hline
\hline
All(\textbf{A1})
&--&--&--&--&
&$-5.27^{+0.00}_{-0.08}$ & $0.0570^{+0.0001}_{-0.0001}$ & $0.5429^{+0.0100}_{-0.0055}$ & $1.1$ & $5.907^{+0.000}_{-0.039}$&$-i\,0.680^{+0.000}_{-0.015}$ \\
All(\textbf{A2})
&--&--&--&--&
&$-5.25^{+0.01}_{-0.05}$ & $0.0570^{+0.0001}_{-0.0001}$ & $0.5385^{+0.0101}_{-0.0057}$ & $1.7$ & $5.919^{+0.002}_{-0.023}$&$-i\,0.684^{+0.002}_{-0.008}$ \\
All(\textbf{A3})
&--&--&--&--&
&$-5.28^{+0.08}_{-0.17}$ & $0.0570^{+0.0001}_{-0.0001}$ & $0.5434^{+0.0102}_{-0.0070}$ & $0.8$ & $5.903^{+0.040}_{-0.085}$&$-i\,0.679^{+0.014}_{-0.032}$ \\
\hline
  \end{tabular}
\caption{\label{tab:IAMfits}
Compilation of all analyses of the lattice configurations as denoted in the first column.
Fits are performed either using Breit-Wigner (BW) or Inverse Amplitude Method (IAM),
on various sets of energy eigenvalues obtained from Lattice using
\textbf{A1}, \textbf{A2} and \textbf{A3}, see main text for more detail on
each of methods and explicit parametrizations. Two light pion mass Breit-Wigner fits yielded a best fit parameter value at the fitting limit, corresponding results are marked by a $\star$. The pole positions of the $\rho$-meson in the complex plane at corresponding pion mass are denoted by $E_\rho$. The last three rows show the results of the combined fit to all lattice data using IAM and pole positions at the lightest (near physical) pion mass.
}
\end{table*}

\section{Extrapolation to the physical point}

The IAM-based global fit strategy for extracting complex pole positions allows
for an extrapolation to the physical point. We define the latter
consistently using the current FLAG~\cite{Aoki:2019cca} values as
$M_\pi^\mathrm{phys}=135$~MeV and
$f_\pi^\mathrm{phys}=130.41$~MeV. Note that since $\cot \delta_1^{\rm
  IAM}$ is parameterized by $\bar l_1-\bar l_2$ and $M_\pi/f_0$, the
only unknown quantity required to study the extrapolation to the
physical point is $M_\pi^\mathrm{phys}$ in lattice units. We found
that the most consistent way to fix the latter is to solve
\begin{align}
\frac{F_\pi^\mathrm{phys}}{M_\pi^\mathrm{phys}}=
 \frac{a f_0}{a M_\pi^\mathrm{phys}}
\left(1 - 2\left(\frac{aM_\pi^\mathrm{phys}}{4\pi af_0}\right)^2
\log\left(\frac{(aM_\pi)^2}{(a\Lambda_4)^2}\right)
\right)\,,
\end{align}
for $aM_\pi^\mathrm{phys}$ with fitted $af_0$ and $a\Lambda_4$ as
input. $a$ is then obtained by setting
$M_\pi^\mathrm{phys}=135$~MeV. Note that this relation is valid up to
the next-to-leading chiral order.
We perform this for each of the
bootstrap samples and methods {\bf A1-3} separately.
Besides phase-shifts and pole positions at the
physical point, this also allows one to extract $f_0$ in physical
units as well as the low-energy constant of interest
$\bar l_4=2\log(a\Lambda_4/(aM_\pi^\mathrm{phys})$.

\section{Fits above inelastic thresholds}

Depending on the analysis method (\textbf{A1-A3}), the spectrum obtained from the $cA2.09.48$ ensemble contains up to 30 energy eigenvalues.
The value of having data close to the physical pion mass is clear, however, only $\sim 5$ energy eigenvalues lie below the inelastic ($4M_\pi$) threshold.
To avoid additional bias only these data points are used in the main part of this letter.

For completeness, we also attempted to fit all the existing data using
the two-body Lüscher formalism, simply to see whether or not $4\pi$ and $6\pi$ interactions contribute a sizable amount to the isovector $\pi\pi$ interaction.
It is known from ChPT that four-particle effects are highly suppressed~\cite{Gasser:1990bv}, which might justify such an approach.

In addition, while approximate, this analysis can provide an estimate of further systematic effects, beyond what is discussed in the main text.

\begin{figure*}
  \includegraphics[width=0.98\textwidth]{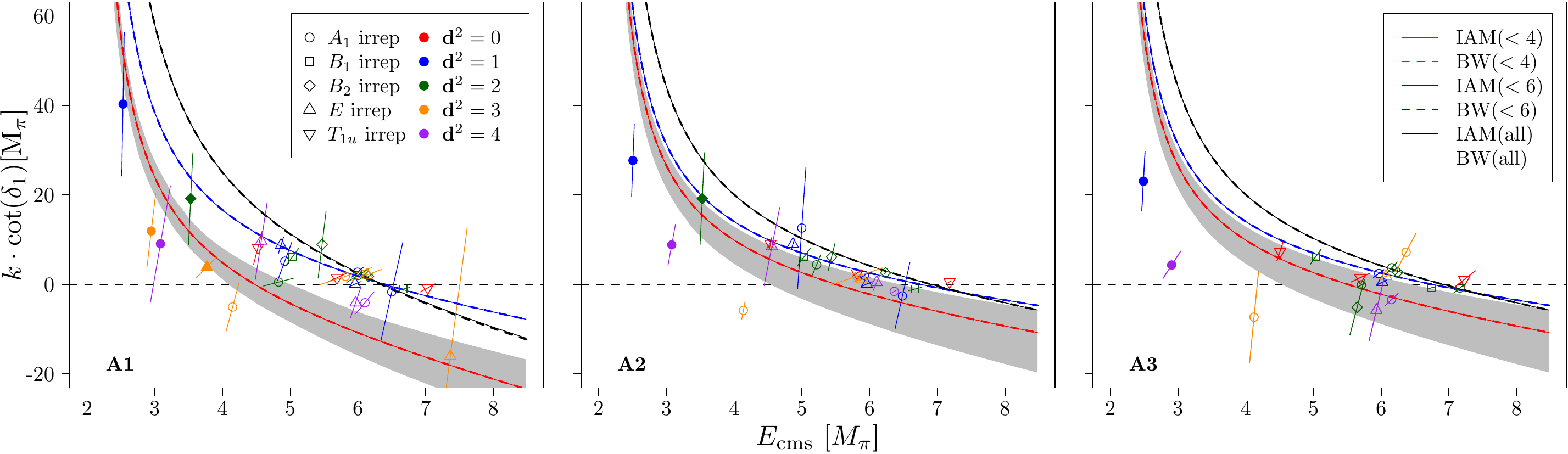}
  \caption{\label{plot:NoLuescher}
    $k\cot(\delta_1)$ as a function of $E_\mathrm{cms}$, both in units
    of $M_\pi$,
    obtained from the fits to energy eigenvalues for the ensemble cA2.09.48
    using the three different analysis (from left to right: \textbf{A1}, \textbf{A2} and \textbf{A3})
    for three different fit-intervals. Full and dashed lines
    show results of IAM- and BW-based fits, respectively.
    Data points are added for illustration purposes only, and are not used in the fits.
  }
\end{figure*}

To avoid mixing with other systematic effects, we perform such exploratory fits separately for each of the used parametrizations (BW and IAM) for each of the methods \textbf{A1-A3}.
To this end, we either include all energy eigenvalues or restrict to regions below $6M_\pi$ or $4M_\pi$, respectively.
The results of these fits are depicted in \cref{plot:NoLuescher}.
In reassuring agreement between all cases we find that fits of data above the $6\pi$ threshold lead to very poor correlated $\chi^2$ values, possibly hinting at sizable effects due to the $6\pi$ channels, either on the lattice (e.g., operator basis size), or the analysis part (e.g. 6-body quantization condition).
On the other hand, fits up to the $6\pi$ threshold converge to values of correlated $\chi^2_{\rm dof}\lesssim2$ for both IAM and BW parametrizations.
This suggests that the effects due to $4\pi$ interactions might be indeed suppressed in this channel.

While further interpretation of these observations needs a more careful study, we point out that all fits lead to similar complex pole positions of the $\rho$-meson of $E_\rho\approx(827-86i,856-125i,821-92i)$~MeV, with the three values corresponding to the three methods \textbf{A1-A3}.

\end{appendix}

\end{document}